========================== cut here =======================================
\documentstyle[12pt]{article}

\begin{document}

\title{Reply to the Comment on "$\Delta I=4$ bifurcation in ground bands of 
even-even nuclei and the Interacting Boson Model"}
\author{H. Toki and L.A. Wu  \\
Research Center for Nuclear Physics, Osaka University, \\
Ibaraki, Osaka 567-0047 , Japan\\
}
\date{May 17, 1999}
\maketitle

Kuyucak and Stuchbery (KS) comment on our study on the
$\Delta I = 4$ bifurcation phenomena in the ground rotational
band \cite{tok} from both the experimental and the 
theoretical sides \cite{kuy}.
Although there exist some interesting systematics in the
existing experimental data, which are
discussed in detail in our second paper \cite{wu},
we have to wait more precise experiments for the definite 
evidence of these interesting phenomena.

The main aim of our study \cite{tok,wu} was to show that there exists 
the mechanism to provide the 
$\Delta I=4$ bifurcation phenomenon in the simplest version of IBA, 
the IBM-1.  The existence of the staggering pattern depends only on 
the geometrical features of 
IBM, the boson number $N$ being finite and $\beta > \sqrt{2}$.
The large value for $\beta > \sqrt{2}$ was questioned 
in the comment \cite{kuy}.  The large $\beta$ value was used,
however, by various authors in the literature 
\cite{gin,war,iac}.

The consequence of the use of the actual value for $\beta$
by Toki and Wu \cite{tok,wu} on the beta and the gamma bands is 
discussed by Kuy et al.
using the quadrupole Hamiltonian \cite{kuy}.  We should, however,
be aware of the fact that
there exist many
Hamiltonians which can give the same value for $\beta$, but 
do not provide the same spectra. 

In fact, our Hamiltonian (used in Fig. 6 of \cite{wu}) 
is in very good agreement with the experimental
systematics not only for the ground band but also for the
beta and the gamma bands.
The reasonable yrast spectra and the suitable amplitude for
the staggering are obtained as seen in Fig. 6 of ref. \cite
{wu}.  We have performed the full diagonalization
calculations using the computor code (PHINT) also for
the beta and the gamma bands.  We get very resaonable 
values for $R_\beta$ value, which ranges from 0.76 to 1.53, and 
$R_\gamma$ from
1.12 to 1.98 for boson number $N$ changed from 4 to 14. This is 
almost in perfect agreement with the
experimental systematics as seen 
in Fig.1 of Ref. \cite{kuy}. These values are much 
better than those of the SU(3) limit, which gives $R_\beta$ and $R_\gamma$ 
approximately 2.5.
 
Concerning the E2 transitions, we have one more 
free parameter $\chi$.  We may take smaller $\chi$ 
in eq. (4) of \cite{kuy} to reduce the value of $R$ 
in Ref. \cite{kuy}.

We want to make one comment on   
the quadrupole Hamiltonian for the case of the large beta
deformation.
The exact diagonalization (PHINT) provides totally
differenct results than the approximate ones using
the $1/N$ expansion \cite{kuy}. 
The amount of the staggering comes out to be 20 keV 
instead of 0.5 keV of the experimental data with $\chi \sim 3.2$
of Ref. \cite{kuy}. 
In order to get the small value for the staggering, one 
should take a much smaller $\chi$, which should be $\sim 1.8$.
However, the $\beta$ and $\gamma$ bandheads are still higher than 
the experimental systematics. 
This was the reason why we did not use the
popular quadrupole formalism in our papers \cite{tok,wu}.


\begin{thebibliography}{99}
\bibitem{tok} H. Toki and L. A. Wu, Phys. Rev. Lett. 79, 2006 (1997).
\bibitem{kuy} S. Kuyucak and A.E. Stuchbery, Phys. Rev. Lett. (1998).
\bibitem{wu} L. A. Wu and H. Toki, Phys. Rev. C56, 1821 (1997).
\bibitem{gin} J. N. Ginocchio and M. W. Kirson, Nucl. Phys. A350, 31 (1980).
\bibitem{war} D. D. Warner and R. F. Casten, Phys. Rev. C28, 1798 (1983).
\bibitem{iac} F. Iachello and A. Arima, The Interacting Boson Model 
(Cambridge, New York, 1987), pp.104-106.

\end{thebibliography}
\end{document}